\documentclass[superscriptaddress,aps,prx,twocolumn,showpacs,nofootinbib, longbibliography, notitlepage,floatfix]{revtex4-2}
\usepackage{etex}
\usepackage{amsmath,amssymb,amsthm}
\usepackage[colorlinks=true,citecolor=blue,urlcolor=blue]{hyperref}
\usepackage[pdftex]{graphicx}
\usepackage{times,txfonts}
\usepackage{braket}
\usepackage{color}
\usepackage{natbib}
\usepackage{amsmath,blkarray}
\usepackage{mathtools}
\usepackage{ulem}
\usepackage{latexsym}
\usepackage{tabularx, booktabs}
\usepackage{graphics,epstopdf}
\usepackage{graphicx}
\usepackage{float}
\usepackage{amsfonts}
\usepackage{tikz}
\usetikzlibrary{quantikz}
\usepackage{color,soul}

\newcommand{\be}{\begin{equation}}
\newcommand{\ee}{\end{equation}}
\newcommand{\ba}{\begin{eqnarray}}
\newcommand{\ea}{\end{eqnarray}}

\usepackage{multirow}
\usepackage{appendix}
\usepackage{url}

\begin{document}
\title{Digitized-Counterdiabatic Quantum Algorithm for Protein Folding} 

\author{Pranav Chandarana}
\affiliation{Department of Physical Chemistry, University of the Basque Country UPV/EHU, Apartado 644, 48080 Bilbao, Spain}
\affiliation{EHU Quantum Center, University of the Basque Country UPV/EHU, Barrio Sarriena, s/n, 48940 Leioa, Biscay, Spain}

\author{Narendra N. Hegade}
\email{narendra.hegade@kipu-quantum.com}
\affiliation{Kipu Quantum, Greifswalderstrasse 226, 10405 Berlin, Germany}
\affiliation{International Center of Quantum Artificial Intelligence for Science and Technology~(QuArtist) \\ and Physics Department, Shanghai University, 200444 Shanghai, China}

\author{Iraitz Montalban}
\affiliation{Kipu Quantum, Greifswalderstrasse 226, 10405 Berlin, Germany}
\affiliation{Department of Physics, University of the Basque Country UPV/EHU, Barrio Sarriena, s/n, 48940 Leioa, Biscay, Spain}

\author{Enrique Solano}
\email{enrique.solano@kipu-quantum.com}
\affiliation{Kipu Quantum, Greifswalderstrasse 226, 10405 Berlin, Germany}
\affiliation{International Center of Quantum Artificial Intelligence for Science and Technology~(QuArtist) \\ and Physics Department, Shanghai University, 200444 Shanghai, China}
\affiliation{IKERBASQUE, Basque Foundation for Science, Plaza Euskadi 5, 48009 Bilbao, Spain}

\author{Xi Chen}
\email{chenxi1979cn@gmail.com}
\affiliation{Department of Physical Chemistry, University of the Basque Country UPV/EHU, Apartado 644, 48080 Bilbao, Spain}
\affiliation{EHU Quantum Center, University of the Basque Country UPV/EHU, Barrio Sarriena, s/n, 48940 Leioa, Biscay, Spain}

\begin{abstract}
We propose a hybrid classical-quantum digitized-counterdiabatic algorithm to tackle the protein folding problem on a tetrahedral lattice. Digitized-counterdiabatic quantum computing is a paradigm developed to compress quantum algorithms via the digitization of the counterdiabatic acceleration of a given adiabatic quantum computation. Finding the lowest energy configuration of the amino acid sequence is an NP-hard optimization problem that plays a prominent role in chemistry, biology, and drug design. We outperform state-of-the-art quantum algorithms using problem-inspired and hardware-efficient variational quantum circuits. We apply our method to proteins with up to 9 amino acids, using up to 17 qubits on quantum hardware. Specifically, we benchmark our quantum algorithm with Quantinuum's trapped ions, Google's and IBM's superconducting circuits, obtaining high success probabilities with low-depth circuits as required in the NISQ era.
\end{abstract}

\maketitle

\section*{Introduction}
Variational quantum algorithms (VQAs) have been proposed to solve problems with noisy qubits in near-term quantum computers~\cite{RevModPhys.94.015004}. VQAs are hybrid classical-quantum algorithms that optimize a cost function containing information about the solution. The quantum part of a VQA consists of a parameterized quantum circuit (PQC), also known as circuit ansatz, to produce trial quantum states. The classical part consists of an optimization routine that gives optimal parameters to solve the problem. The choice of PQC affects the performance of the VQA to a great extent. These PQCs are broadly divided into two categories: problem-inspired and hardware-efficient. Problem-inspired ansatz utilizes the properties of the problem Hamiltonian to efficiently reach the expected state. Hardware-efficient ansatz takes the information of the device connections to reduce the noise due to deep circuits and unimplementable connections. Some examples of problem-inspired ansatz are the Quantum Approximate Optimization Algorithm (QAOA) ~\cite{farhi2014quantum}, the unitary coupled-cluster ansatz~\cite{Romero_2018} or the Hamiltonian variational ansatz~\cite{PhysRevA.92.042303,PRXQuantum.1.020319}. On the other hand, some noteworthy hardware-efficient ans\"atze can be found in   Refs.~\cite{Robert2021,kandala2017hardware}.

Implementing VQAs is challenging due to shot noise, measurement noise, and others. It has also been shown that VQAs suffer from barren plateaus, where the gradients vanish with increasing system size~\cite{McClean2018,PRXQuantum.3.010313, Larocca2022diagnosingbarren}. Generally, VQAs with hardware-efficient ans\"atze suffer from this challenge due to their high expressibility at larger depths. Hence, the use of the problem-inspired ansatz is motivated. In problem-inspired ansatz, the limited search space results in lower expressibility and higher trainability. That being said, problem-inspired ans\"atze usually involve large circuit depths, so experimental implementation becomes unfeasible on available noisy devices. Thus, a good circuit ansatz has to be expressible so that it contains the solution, but not too expressible that it becomes untrainable.  

Recently, several works have reported the use of digitized-counterdiabatic quantum computation (DCQC) to improve and compress state-of-the-art quantum algorithms.  These methods utilize counterdiabatic (CD) protocols to accelerate given adiabatic quantum algorithms to generate many-body ground states~\cite{hegade2021shortcuts} and QAOA~\cite{chandarana2022digitized}. Furthermore, they have also shown drastic improvements in industrial applications, like portfolio optimization~\cite{hegade2021portfolio} and integer factorization~\cite{PhysRevA.104.L050403}. These methods have some difficulties, like finding suitable initial parameters and optimal CD terms. To solve these challenges, a meta-learning technique was proposed recently to find suitable initial parameters~\cite{chandarana2022meta}. The choice of optimal CD terms may be tackled by machine learning methods like reinforcement learning~\cite{yao2021reinforcement} and Monte-carlo tree search~\cite{yao2022monte}. CD protocols stem from the field of shortcuts to adiabaticity, which was developed to accelerate the quantum adiabatic processes. Among many methods, like fast forward~\cite{masuda2008fast,masuda2010fast} and invariant-based engineering~\cite{chen2010fast,chen2011lewis}, CD driving~\cite{demirplak2003adiabatic,demirplak2005assisted,berry2009transitionless} has been of prominent interest over the years for studying many-body quantum systems. Apart from shortcuts to adiabaticity, other quantum control protocols like quantum optimal control (QOC) are also studied in the context of VQAs~\cite{PRXQuantum.2.010101,meitei2020gate}. VQAs have also been implemented to find optimal control sequences~\cite{PhysRevLett.118.150503}.

In this article, we develop a hybrid classical-quantum digitized-counterdiabatic algorithm to tackle a protein folding problem. It consists of a PQC inspired by CD protocols and a classical optimization routine for parameter optimization. Proteins are macromolecules consisting of a large chain of amino acid residues and perform many vital functions in organisms like DNA replication, catalyzing metabolic reactions, and more. Knowledge of how proteins fold is crucial in understanding enzymes. The mechanics of folding may unravel remedies for diseases like Alzheimer’s, Huntington’s, and Parkinson’s that are induced due to misfolding of proteins. With the combinatorially increasing solution space, protein folding problem is highly complex for classical computation, which suggests the potential use of quantum computers. 

Generally, protein folding is modeled by suitable 2D or 3D lattices, while the amino acids are allowed to be placed such that the interaction energy is minimized. By proper encoding schemes, this problem can be converted into a problem Hamiltonian whose ground state shows the configuration of the concerned protein in the given lattice. Over the last decade, numerous attempts have been made to tackle this problem with quantum computing. For instance, Perdomo et al.~\cite{PhysRevA.78.012320} studied a hydrophobic polar (HP) model with a 2D lattice and later also studied a coarse-grained model (3D lattice) with quantum annealing ~\cite{Perdomo-Ortiz2012}. Babbush et al.~\cite{doi:https://doi.org/10.1002/9781118755815.ch05} also studied the protein folding problem with turn encoding in quantum adiabatic algorithms. Babej et al.~\cite{babej2018coarse} studied this problem using the quantum alternating operator ansatz~\cite{a12020034}. Recently, a resource-efficient version of the same problem was considered by Anton Robert et al.~\cite{Robert2021} with an experimental demonstration on IBM superconducting device. 

To address this problem, we propose a hybrid digitized-counterdiabatic algorithm that includes a PQC we call ``CD-inspired" ansatz. While being problem-inspired, this ansatz is also hardware implementable and has a parameterization that scales as $\mathcal{O}(N^2)$, where $N$ is the number of qubits. We benchmark the performance against state-of-the-art problem-inspired ansatz as well as hardware-efficient ansatz. In addition, we also perform experiments with system sizes up to 17 qubits using several quantum hardware platforms with different connectivity and native gates.
\begin{figure*}
    \centering
    \includegraphics[width=1\linewidth]{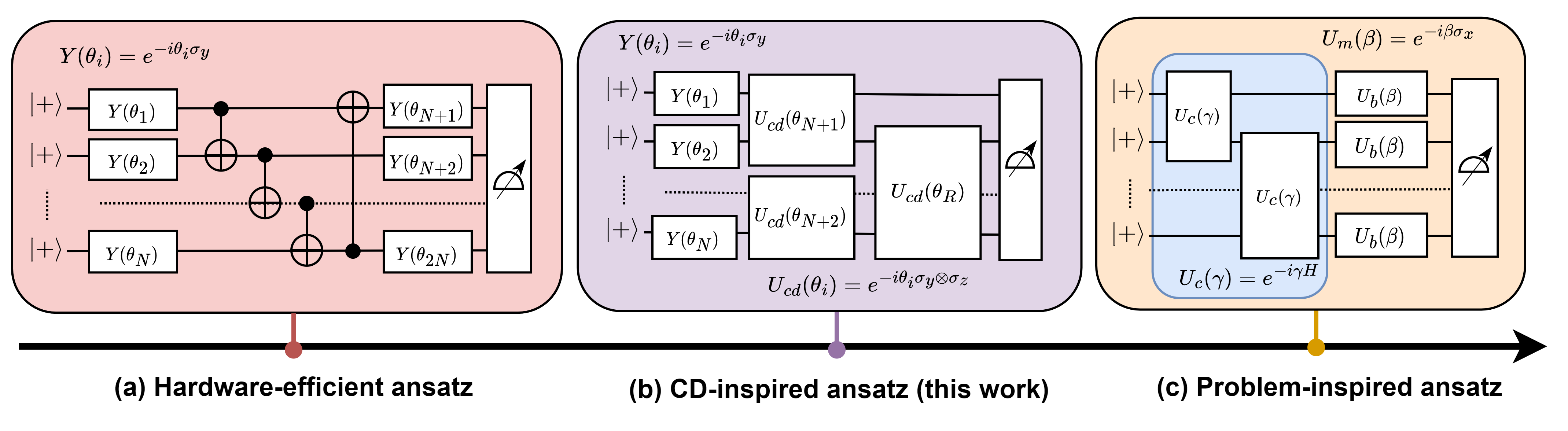}
    \caption{A Schematic diagram of different types of ansatz with $p=1$ layer. In (a) hardware-efficient ansatz (HEA), which has no information from the problem, (b) CD-inspired ansatz proposed in this work, and (c) QAOA, which is a problem-inspired ansatz. Horizontal lines show qubit registers that are initialized in $\ket{+}^{\otimes N}$ state. In HEA, there are parameterized $Y$ rotations followed by nearest-neighbor entangling gates, then again parameterized $Y$ rotations. In QAOA, we have Hamiltonian term $U_c(\gamma)$ and mixer term $U_b(\beta)$. And, for the CD-inspired ansatz, we have parameterized $Y$ rotations followed by $YZ(\theta)$ rotations. In all cases, cost function $C$ (Eq.~\eqref{exp_ham}) is computed, then the parameters are updated using gradient-based classical optimizers until $C$ is minimized. As we go from left to right, the implementation difficulty increases. }
    \label{fig:Schematic-PF-circuit}
\end{figure*}

\section*{Results}
\textbf{CD-inspired ansatz.} In this section, we will discuss the construction of the CD-inspired ansatz for the algorithm and study its performance. We begin by considering quantum adiabatic evolution with counterdiabatic protocol Hamiltonian $H_{cd}$ given by
\begin{equation}
    H_{cd}(t) = ( 1 - \lambda(t))H_{mixer} + \lambda(t)H + \Dot{\lambda}(t) A_\lambda\label{cdevol},
\end{equation}
where $H_{mixer}$ is a Hamiltonian whose ground state is easy to prepare, $\lambda(t)$ is a scheduling function with boundary conditions $\lambda(0) = 0$, $\lambda(T)=1$, $T$ is the total evolution time, and $A_{\lambda}$ is the approximate CD term, calculated by using the nested commutator (NC) method (See Methods~\ref{Variational quantum algorithms and counterdiabaticity}). Lower orders of $A_\lambda$ give the approximate CD terms while $l \to \infty$ will give the exact CD term.
While working in the adiabatic regime, the scheduling function $\lambda(t)$ should be slow enough to satisfy the adiabatic theorem to reach the ground state of the target Hamiltonian with high probability. However, with the CD term, this condition is lifted as the non-adiabatic transitions can be suppressed~\cite{sels2017minimizing} and at $|~\Dot{\lambda}~| \to 0$, we retrieve the adiabatic Hamiltonian. Now consider a certain $\lambda(t)$ that satisfies $|~\Dot{\lambda}(t)~| \gg |~\lambda(t)~|$ . For this scenario, the evolution will almost be non-adiabatic and most of the contribution will be from $A_\lambda$ term. In theory, this evolution should also be successful but it will require the calculation of the exact CD term which is a challenging task as information on all the spectral properties of the Hamiltonian becomes necessary. 

In DC-QAOA~\cite{chandarana2022digitized}, the Eq.~\eqref{cdevol} is digitized with approximate CD terms, to get faster evolution. Instead of using actual scheduling functions, we take the aid of classical optimization routines to optimize trotter evolution to reach to the ground state. Under the assumption that there exists a scenario as mentioned above, we can get rid of the first two terms of Eq.~\eqref{cdevol} and only the CD terms can be considered. This is the intuitive motivation behind the CD-inspired ansatz: Instead of implementing all the evolution, we implement only the contributions from the digitized CD term as a parameterized circuit and allow the classical optimization to take care of the evolution to lead to the ground state. Thus, the CD-inspired ansatz will have the form
\begin{equation}
    U_{cd}(\theta) = e^{-i \theta A}\label{cdeq},
\end{equation}
where $A \in A_\lambda$ and $A_\lambda$ is a set of all the terms computed from the NC method. This is advantageous for VQAs in the sense that this condition gets rid of most of the terms from the ansatz which makes it implementable in the near-term devices and as these algorithms aim to find approximate solutions, this ansatz should lead to good solutions if a suitable optimization strategy is used. This claim can also be backed by QOC theory where the dynamical Lie algebra of non-commuting control Hamiltonians can be computed in order to understand the performance of the ansatz~\cite{anand2022exploring}. However, as the terms in the CD pool operators, $A_\lambda$ are calculated both by executing only odd commutations (See Methods~\ref{Variational quantum algorithms and counterdiabaticity}), it will yield a relatively compact pool of operators. Thus, the NC method provides us with a way to truncate the full expansion and then select the appropriate terms from the reduced pool of operators.  

Apart from that, an important task is to choose the parameterization of the ansatz. There are several ways to accomplish this, for instance by studying the symmetries of the problem Hamiltonian~\cite{sauvage2022building}. Given we are relying on few terms from the NC method, it is beneficial that each term has its own free parameter, increasing the degrees of freedom of the ansatz.

Summing up, the algorithm will consist of a PQC initialized in the ground state of $H_{mixer}$. Then we choose a pool of operators obtained from Eq.~\eqref{cdeq} to construct the ansatz, where each term has its free parameter to be optimized by a classical optimizer. The number of free parameters will depend on the number of interaction terms in the Hamiltonian. A schematic diagram of hardware-efficient ansatz, CD-inspired ansatz, and problem-inspired ansatz is shown in Fig.~\ref{fig:Schematic-PF-circuit}.

\begin{figure*}[t]
    \centering
    \includegraphics[width=1\linewidth]{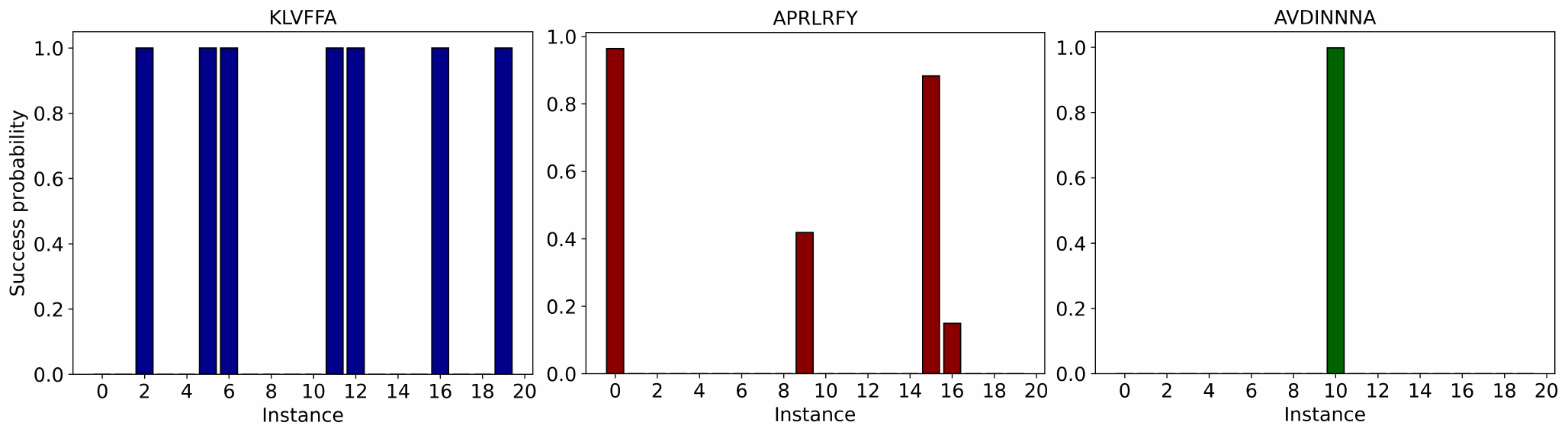}
    \caption{Success Probability as a function of 20 randomly initialized instances for $6$, $7$, and $8$ amino acid proteins with $N= 6$, $N=9$, and $N=13$ qubits respectively using the CD-inspired ansatz with $p=1$ layer, maximum 500 iteration steps and gradient-based optimizers Adam and Adagrad were used for classical optimizations. }
    \label{fig:SPvsIte}
\end{figure*}

\textbf{Performance analysis.} In this section, we study the application of the CD-inspired ansatz to various proteins with different numbers of amino acids. These include the amyloid-beta peptide sequence (KLVFFA) which translates to a 6-qubit system, the Neuropeptide-alpha bag cell (APRLRFY), which translates to a 9-qubit system, cyclic peptide inhibitor (AVDINNNA)    which translates to a 13-qubit system and Oxytocin (CYIQNCPLG) which translates to a 17-qubit system. Each of the letters represents an amino acid, for example, A: Alanine, C: Cysteine, D: Aspartic Acid, among others, with the Miyazawa and Jernigan (MJ) interactions~\cite{MIYAZAWA1996623}.

In each case, we generate a 5-local Ising Hamiltonian (See Methods~\ref{Protein folding}) intending to reach the ground-state, that shows the protein configuration by minimizing the expectation value. By considering the NC commutator in Eq.~\eqref{gauge} with $l=2$, we obtain a set of operators with increasing locality. We truncate these terms to up to two-body terms, which result in $U_{cd}=\{Y, YZ, ZY, XY, YX\}$. Here, each of the terms shows the exponentiation of the corresponding Pauli terms, for example $XY= e^{-i \sum_{i,j} J_{ij} \sigma_i^x \sigma_j^y}$.  Out of these, we select $\{Y, YZ\}$ where $Y= e^{-i \sum_m  \sigma_m^y}$ with $m=0,..., N$, and $YZ= e^{-i \sum_{i,j} J_{ij} \sigma_i^y \sigma_j^z}$ where $(i,j)$ correspond to two-body interaction sites of the Hamiltonian with coefficients $J_{ij}$ (Fig.~\ref{fig:Schematic-PF-circuit}(b)). The number of two-body terms $N_{2loc} \leq N(N-1)/2$ make this ansatz hardware-efficient, in the sense that its experimental implementation is much more feasible. Regarding the optimizable parameters, each of the gates that are applied has its free parameter. Hence, the number of parameters per layer is $R= N_{2loc} + N$ so the parameter scaling is $\mathcal{O}(N^2)$. In Appendix~\ref{parameterscale}, we have shown the parameterization as a function of the system size for various ansatz.

For the classical optimization part, we implement stochastic gradient-descent-based optimizers, called Adam~\cite{kingma2014adam} and Adagrad~\cite{adagradoriginal}, where the gradients are computed using the parameter-shift rule~\cite{PhysRevA.99.032331}. For each protein, we run the algorithm 20 times with random initial parameters for the $p=1$ layer. To quantify the performance of the algorithm, we use success probability as a metric, showing the probability of getting the ground state at the end of the algorithm. We also study the expectation values as a function of the iteration steps to understand the convergence. 
\begin{figure}
    \centering
    \includegraphics[width=1\linewidth]{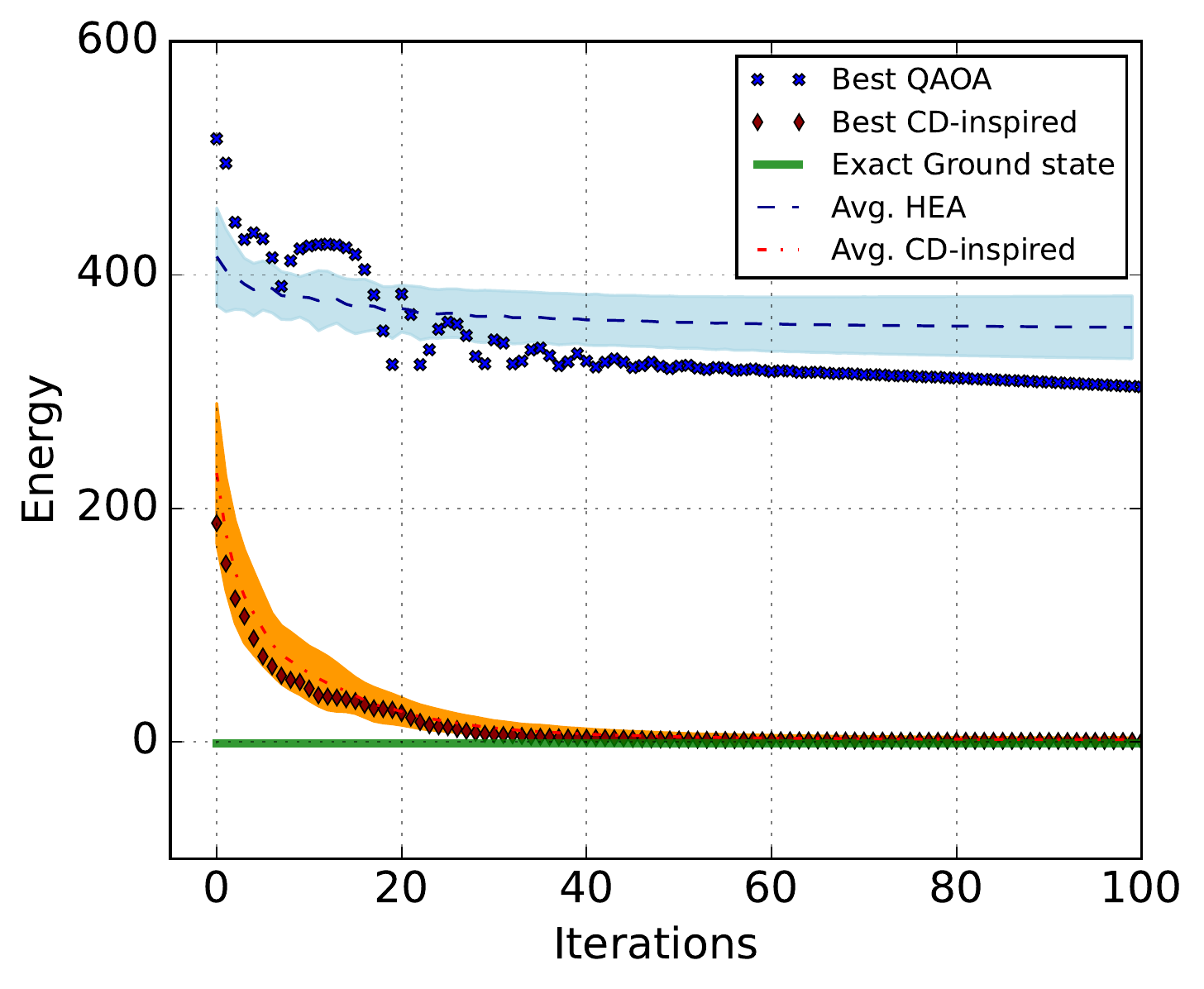}
    \caption{Energy as a function of iterations for $N=13$ qubit protein AVDINNNA. Results show simulator data for 100 iterations using Adam optimizer comparing QAOA and CD-inspired ansatz. Dashed lines display the average energy of 20 instances considered. Crosses show the energy variation of the best instance. Shaded regions indicate the standard deviation. The green line shows the exact ground state energy.}
    \label{fig:QAOAvsCA}
\end{figure}

To benchmark the performance, we implemented $p=1$ algorithm as a function of the number of instances. For each of them, the ansatz was randomly initialized and the success probabilities are shown in Fig.~\ref{fig:SPvsIte}. The ansatz performs extremely well for all the cases we investigate but with increasing system size, the number of successful instances decreases. This is due to the fact that the solution space increases drastically with system size making it difficult to find a global solution. For the numerical simulations, we set the number of iterations to 500 with the tolerance of $10^{-6}$,  in the sense that the iterations will stop once this tolerance is reached. The `unsuccessful' instances show almost zero success probability, this behavior can be attributed to the fact that the spectrum of energy eigenstates is densely packed and has many degenerate excited states, so there is a finite possibility of the output state being stuck at the lower excited states with high probability. These results were computed with the intent to check how efficiently the algorithm can reach the exact ground state. However, the hybrid classical-quantum algorithms usually aim at finding approximate solutions near the ground state. The observed trend of decreasing successful instances with increased system size shows that scaling to a higher systems size might require good initialization techniques to get to the exact ground state. 

To evaluate CD-inspired ansatz against state-of-the-art quantum algorithms, we begin by doing a performance comparison with a well-known problem-inspired ansatz: QAOA. To do so, we studied the convergence of 20 instances with $p=1$ layer, each having random initialization, with $N=13$ qubits protein AVDINNNA  and the results are shown in Fig.~\ref{fig:QAOAvsCA}. Average energy with standard deviation and the best instance (in terms of convergence) is plotted as a function of iteration steps and compared with the exact ground state energy. Indeed, the performance of the CD-inspired ansatz is much better than QAOA. This performance enhancement can be attributed to the difference in the optimizable parameters for both algorithms. Only two parameters result in low expressibility, and to counter this challenge, a high $p$ ansatz is required. Taking into consideration that QAOA ansatz at $p=1$ already contains five-body terms, increasing layers would serve to be problematic as the two-qubit gate errors will start to accumulate and the circuit depth will also increase, leading to decoherence. Apart from this, the standard deviation of the average energy of CD-inspired ansatz is initially high but it reduces as the iteration steps increase. Hence, the algorithm will lead to states close to the ground state independent of the initial parameters chosen. Lastly, the algorithm reaches the approximate solutions in a relatively low number of iteration steps. This behavior is useful in the hardware setting as it will reduce the run-time of the algorithm.

\begin{figure}[]
    \centering
    \includegraphics[width=1\linewidth]{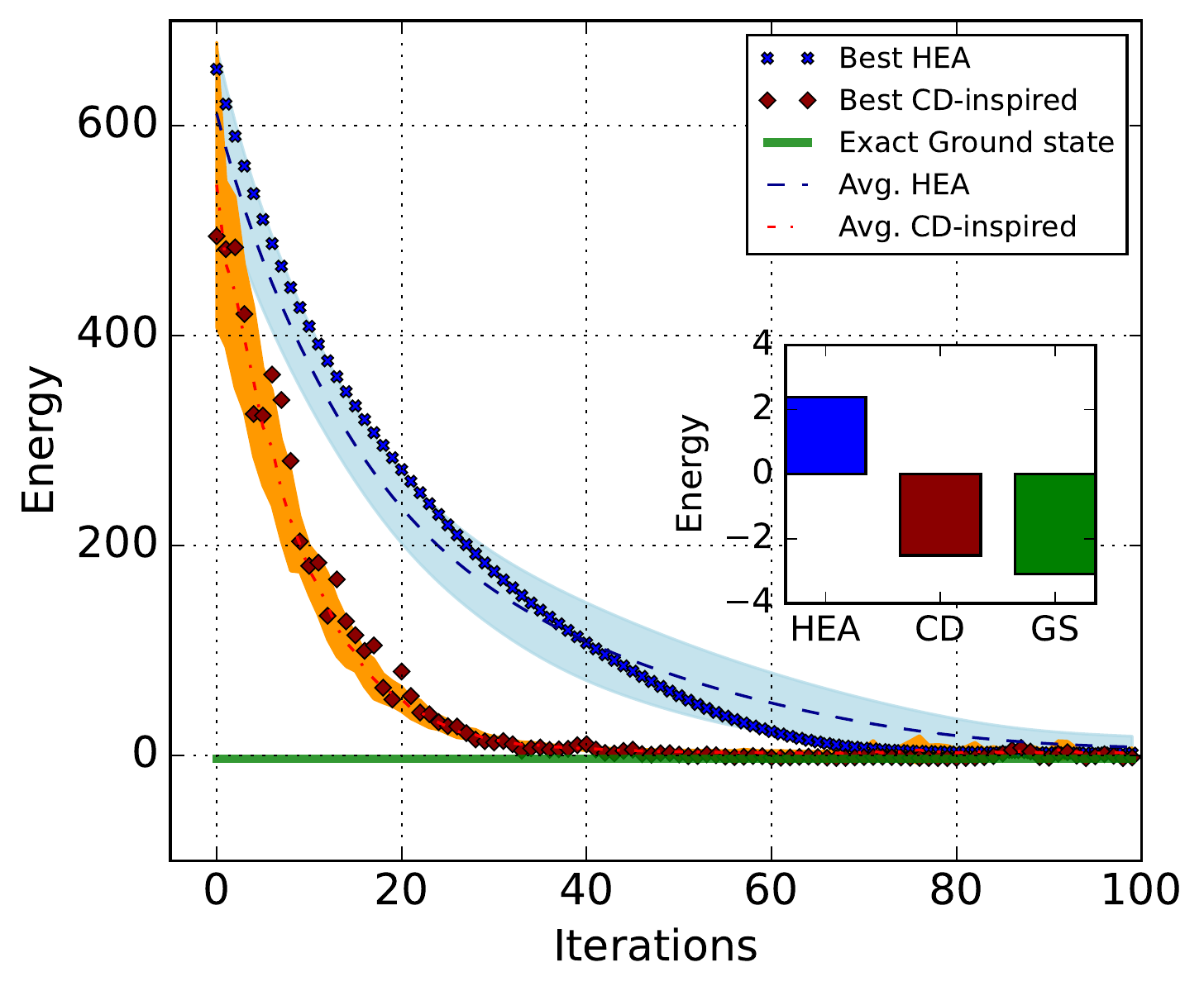}
    \caption{Energy as a function of iterations for $N=17$ qubits system comparing CD-inspired ansatz and hardware-efficient ansatz. Numerical simulations were performed for 100 iterations with the Adam optimizer. Dashed lines show the average energy of the 20 instances considered and Crosses show the best instance. The inset plot shows the minimum energy achieved during the optimization of the best instance. The green line shows the exact ground state.   }
    \label{fig:new_CYQNCPLG_HEA}
\end{figure}

After comparing to a problem-inspired ansatz, we compare CD-inspired ansatz with a hardware-efficient ansatz (HEA). To do so, we implemented a circuit with parameterized $Y$ rotation applied to all qubits followed by CNOTs to the cyclic nearest neighbors for entanglement, followed by parameterized $Y$ rotation applied to all qubits (Fig.~\ref{fig:Schematic-PF-circuit}(a)). We study a nine amino-acid protein CYIQNCPLG that translates to a $N=17$ qubits system by comparing both average energy and best instance for $p=1$ algorithm and results are shown in Fig.~\ref{fig:new_CYQNCPLG_HEA}. The inset plot shows the minimum energy obtained during the optimization of the best instance. We observe that the minimum with the best CD-inspired ansatz instance is much closer to the exact ground-state energy as compared to HEA. Regarding the convergence with iteration steps, the HEA has a relatively smoother convergence as compared to the CD-inspired ansatz. As far as the circuit evaluations are concerned, the parameter-shift method will require 2 evaluations per parameter to estimate the gradients, hence the CD-inspired ansatz will take $2(N_{2loc} + N)$  while HEA will take $2N$ circuit evaluations. This is a challenge since, at larger system sizes, these would increase drastically. To circumvent this, different methods to compute the gradients can be adopted. For instance, Ref.~\cite{hoffmann2022gradient} studies a method that can compute the gradients with only two circuit evaluations; independent of the number of circuit parameters.  Besides this, the energy difference between the ground state energy $E_{GS}$ and the minimum energy obtained from the best CD-inspired ansatz instance $E_{cd}$ is $E_{cd}-E_{GS} \approx 0.54$. Thus, even at a large system size, we get solutions very close to the ground state. With the CD-inspired ansatz, we observed that for a system size as high as $N=17$ qubits, the ansatz finds it hard to converge to some steady value at 100 iterations which might be due to the classical optimization routines or lower iteration steps. This also means that near the ground state, the energy landscape is extremely featured which makes it harder for the local optimizer to converge to a particular energy value. 
\begin{figure}
    \centering
    \includegraphics[width=1\linewidth]{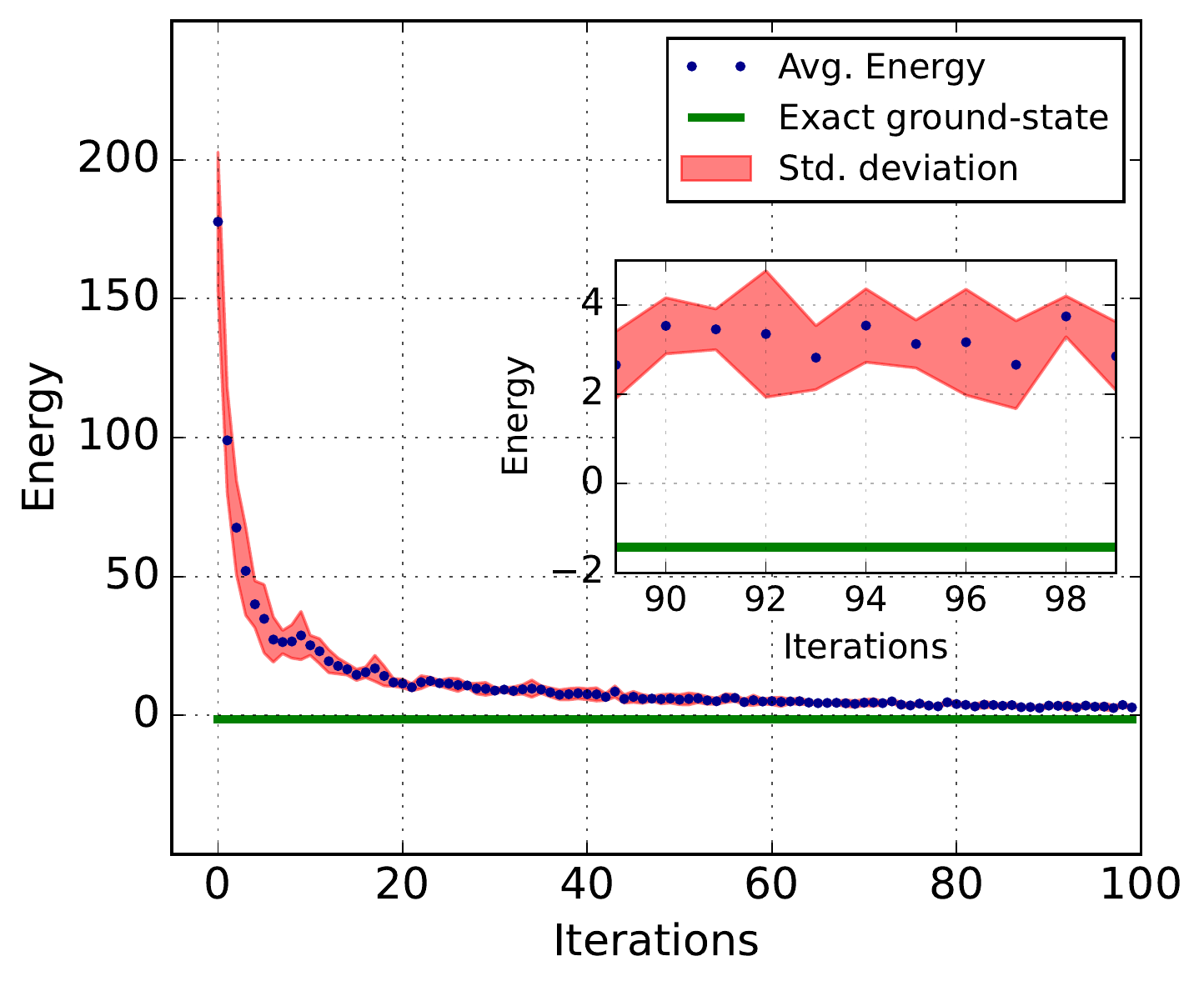}
    \caption{Energy as a function of iterations for $N=9$ qubits with noise model for \textit{ibmq\_guadalupe}. Numerical simulations were performed for 100 iterations with Adagrad optimizer for 5 instances. Blue dots show the average of these instances and the red-shaded region shows the standard deviation. The inset shows the last 10 iteration steps. The green line shows the exact ground state energy.    }
    \label{fig:noisy_sim_9}
\end{figure}

In order to study the performance of the algorithm against noise, we considered a model corresponding to \textit{ibmq\_guadalupe} device from IBM. This model uses the actual backend parameters to mimic noise with some exemptions, namely, the cross-talk errors and leakage errors are not included~\cite{9283531}. Noisy simulations are key to understanding how the energy landscape changes under the effect of noise. In Fig.~\ref{fig:noisy_sim_9}, we take the optimal initial parameters (of ideal simulation) of a $N=9$ qubit system APRLRFY with $p=1$ and perform the algorithm against the noise model five times with Adagrad optimizer and plot the average energy as a function of iterations. The aim is to check how much noise changes the convergence of the optimal algorithm.

We can observe that even in the presence of noise, the ansatz does work considerably well. Having said that, the final convergence energy is lifted considerably as compared to the ground state energy and the algorithm is not exactly converging. There are still fluctuations as we reach the end (refer inset Fig.~\ref{fig:noisy_sim_9}), which is natural due to existing noise. This gives an intuition about the energy landscape when CD-inspired ansatz is implemented. From Fig.~\ref{fig:noisy_sim_9}, it is evident that the energy landscape with CD-inspired ansatz and gradient-based optimizers is smooth enough to take the state near to the ground state but there exist lots of local minima near the exact ground state. These landscape features make it easier for the classical optimizer to get near to the ground state but to reach exactly to the ground state becomes difficult with increasing system size. This fact is also evident from Fig.~\ref{fig:SPvsIte}, as with increasing system size the number of successful instances decreases. Hence, at higher system sizes, there might be a requirement for global optimizations to efficiently reach the ground state with a high success probability. Also, these noisy simulations were run without any error mitigation techniques. The involvement of error mitigation techniques is crucial at the time of experimental implementation. There has been significant attention in studying error mitigation techniques~\cite{PhysRevA.103.042605,ding2020systematic,9259940,PhysRevX.7.021050,PhysRevLett.121.220502,PhysRevLett.119.180509,ravi2022navigating}. Nevertheless, we still are able to reach considerable approximate solutions while performing the algorithm under noise.

With the advantages of implementing this ansatz, several challenges also need to be addressed. First, the choice of suitable CD terms from the pool of NC commutation operators in this work is heuristic but clever techniques need to be developed for choosing the dominant CD term that performs the best depending on the problem. It can also be seen that, when the ansatz does not perform well, the ground state success probability is close to zero. This means that when we scale to higher system sizes, it might be difficult to find the exact ground state as the size of the Hilbert space will be huge. Lastly, although we have successfully reduced the circuit depth, there will still be a requirement for additional circuit optimization strategies to improve the results while performing experiments on real hardware, which we show in the next section for several cloud quantum computing platforms.
\begin{figure*}[t]
    \centering
    \includegraphics[width=1\linewidth]{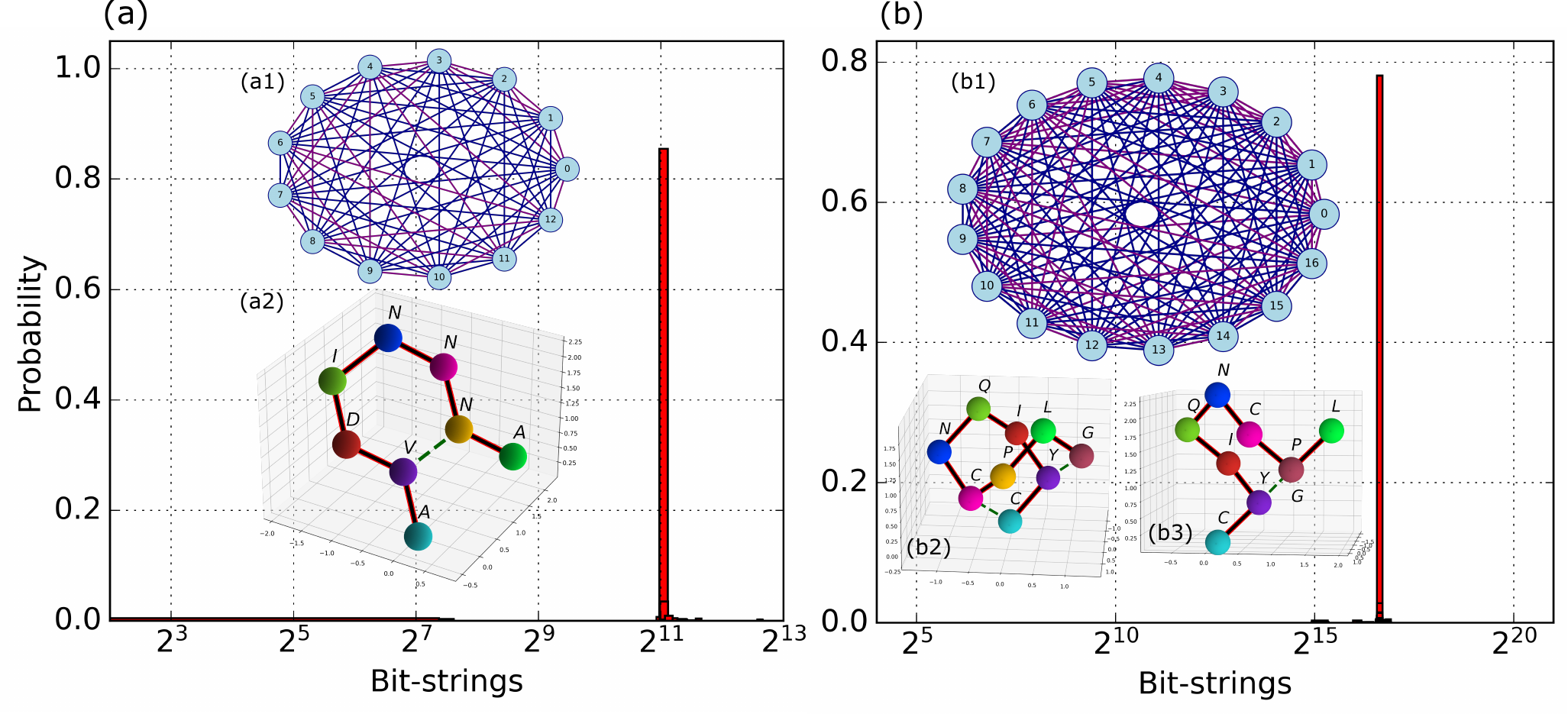}
    \caption{Output probability distribution of $N=13$ AVDINNNA protein and $N=17$ CYIQNCPLG protein on a trapped-ion system: Quantinuum system H1 with 1000 shots. (a) show the $N=13$ case and (b) show the $N=17$ case. (a1) and (b1) show the graph corresponding to two-body interactions implemented in the CD-inspired ansatz. Blue edges show the present two-body connections while the purple edges show the connections that are absent. (a2) shows the optimal protein configuration with a dotted green line depicting the connection of the nearest-neighbor interaction. (b2) show the optimal configuration of protein obtained from exactly solving the problem whereas (b3) shows the protein configuration obtained in the experiment. In (b3) amino acids `P' and `G' are overlapping.   }
    \label{fig:experiment_trapped_ions}
\end{figure*}

\textbf{Experimental implementations.} In this section, we implement the proposed CD-inspired ansatz on different available noisy hardware and emulators, specifically on trapped-ions and superconducting systems. Having various native gate sets and connectivity, all these devices pose different challenges that require to be dealt with to get appreciable results. Details about the hardware, circuit optimization, and error mitigation techniques are given in Appendix~\ref{ibmqappex}.
\begin{figure*}[t]
    \centering
    \includegraphics[width=1\linewidth]{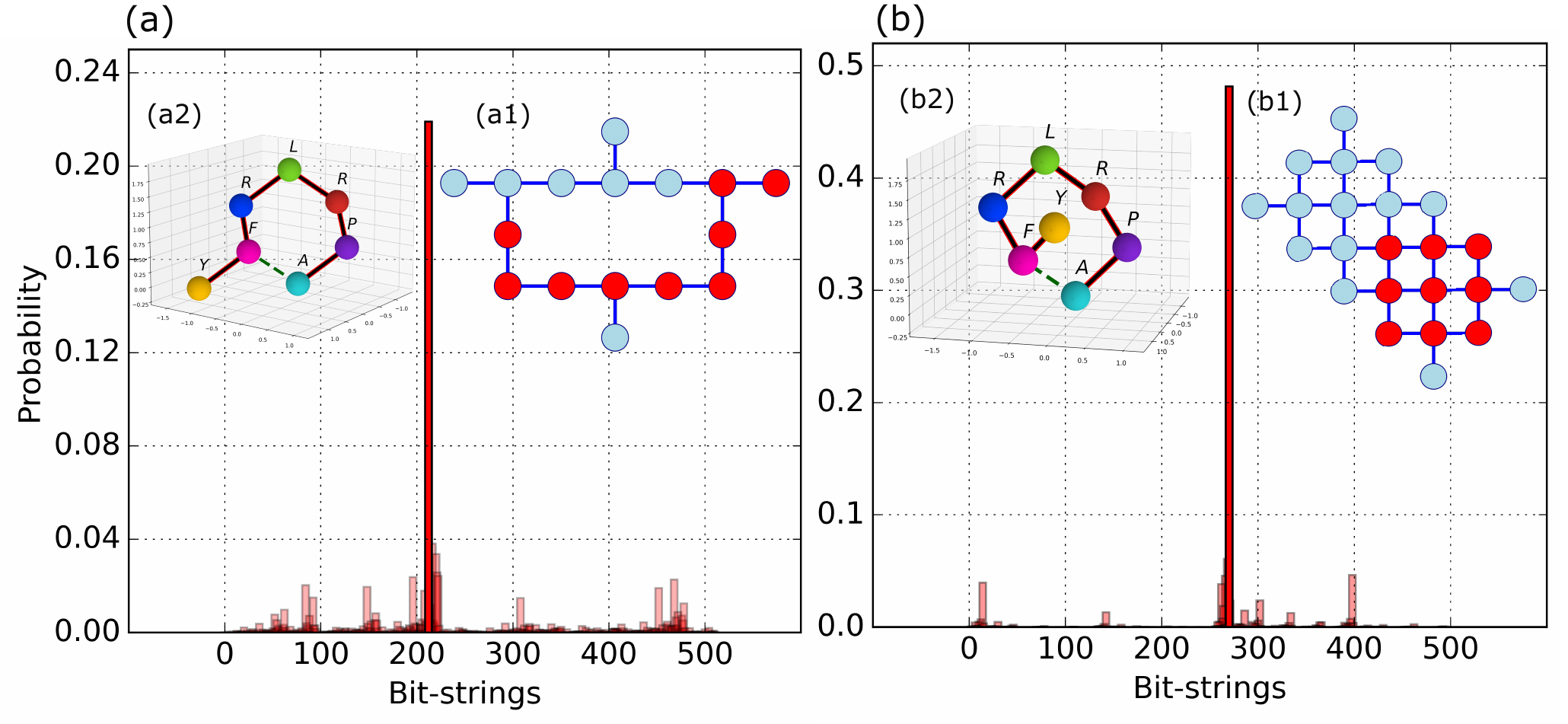}
    \caption{Output probability distribution of $N=9$ qubits by implementing the optimal circuit on (a) IBM $\textit{ibmq\_guadalupe}$ where the experiment was performed with 8192 shots and (b) Google's quantum virtual machine \textit{rainbow}~\cite{isakov2021simulations} where the experiment was performed with 10000 shots. Dark-colored bars show the ground-state probability of the physical qubits whereas light-colored bars show the rest of the distribution  (a1) and (b1) show the hardware topology and selected qubits are shown using red color. (a2) and (b2) both show the optimal protein configurations with the nearest neighbor connection between `A' and `F' shown by a dotted green line. }
    \label{fig:superconducting_expt}
\end{figure*}

\subsubsection{Quantinuum Trapped-ions} We implemented two systems, 8 amino acid protein AVDINNNA ($N=13$ qubits) and 9 amino acid protein CYIQNCPLG ($N=17$ qubits) with $p=1$ layers on the Quantinuum H1-1~\cite{quantinuum} device. For the trapped-ions system, all the qubits are identical and errors depend upon the interaction zones, so the selection of qubits becomes trivial since we can choose any. In both cases, the system was initialized in the $\ket{+}^ {\otimes N} $ state by applying Hadamard gates to all the qubits. Following that, we implement parameterized $R_y({\theta_i})$ rotations on all qubits, and the interaction terms $YZ(\theta)$ were constructed by the native $ZZ(\theta)$ interaction. This is done by applying two rotations, $YZ^{ij}(\theta_m) \equiv R_x^i(\pi/2)~ ZZ^{ij}(J_{ij}\theta_m)~ R_x^i(-\pi/2)$ (refer Fig.~\ref{fig:nativegatecirc}(a)). For $N=13$ system, the Hamiltonian we considered had $N_{2loc}=52$ two-body interactions and for system with $N=17$ qubits, $N_{2loc}=80$ two-body interactions were present. Graphs in Fig.~\ref{fig:experiment_trapped_ions}(a1) and Fig.~\ref{fig:experiment_trapped_ions}(b1) show the full connectivity with blue edges showing the present interactions and purple edges showing the interactions that are absent. When dealing with the real hardware, gates are applied in a parallel manner to reduce the circuit depth but this comes at a cost of performing multiple operations at the same time. In this system, there is a limit on how many parallel operations can be performed efficiently, so we apply the gates such that they do not exceed 5 operations at a time. Due to the limited hardware access, we ran the optimization part on a local simulator with Adam optimizer and 500 iteration steps and implemented the circuit obtained with the optimal parameters on the real hardware.

In Fig.~\ref{fig:experiment_trapped_ions}, we show the probability distribution from the real hardware for both systems. Both experiments were performed with 1000 shots. For $N=13$ qubits, we achieve around 85\% probability of getting the ground state shown in Fig.~\ref{fig:experiment_trapped_ions}(a), while Fig.~\ref{fig:experiment_trapped_ions}(a2)  shows the folded protein in 3D. The nearest-neighbor amino acid connection is `V'--`N', which gives us $q_{2,7}=1$. The optimal turn sequence is $t_{expt} = [\bar1, 0, \bar1, 2, \bar0, 1, \bar0]$. On the other hand, Fig.~\ref{fig:experiment_trapped_ions}(b)  shows the $N=17$ qubits case with around 79\% probability of the output state achieved during the experiment which is the 4th excited state of the Hamiltonian. As the system size is large and we are implementing only $p=1$ layered ansatz, the classical optimization routine gets close to the ground state energy but still reaches local minima. Fig.\ref{fig:experiment_trapped_ions}(b2) shows the protein configuration with the ground state, while Fig.\ref{fig:experiment_trapped_ions}(b3) is the protein configuration with the fourth excited state. The optimal turn sequence is $t_{opt} = [\bar1, 0, \bar3, 1, \bar0, 2, \bar1, 3]$ and the turn sequence we get while experimenting is $t_{expt} = [\bar1, 0, \bar3, 2, \bar0, 3, \bar1, 1]$. We can notice that the last turn taken by the amino acid is non-optimal which leads to overlap at the second last lattice point. For optimal configuration, the nearest-neighbor connections are `C'--`C' and `Y'--`G', and hence $q_{1,6}=1$ and $q_{2,9}=1$. However, with the 4th excited state the connections are `Y'--`G' and `Y'--`P', hence $q_{2,9}=1$ and $q_{2,7}=1$. So, for the fourth excited state, there is one connection and several turns that differ from the exact ground state. In terms of the energy difference, both these configurations differ by a factor of 0.54. This is close to the actual ground state and, thus, the algorithm is likely to jump to the excited state. Implementation of our hybrid quantum algorithm results in high success probability. This is due to the all-to-all connectivity and the fact that the two-body interactions in the circuit can be applied effectively with native gates.

\subsubsection{IBM Superconducting chip} We implemented the 7 amino acid protein APRLRFY on 16-qubit $\textit{ibmq\_guadalupe}$~\cite{ibmdev} device with $p=1$ layer of the ansatz. To choose 9 out of the 16 available qubits, we apply a subgraph isomorphism algorithm~\cite{Nation2022} to get the best layout possible. As usual, we start with the $\ket{+}^ {\otimes N} $ state by applying the Hadamard gate to all the qubits and then apply $R_y(\theta_i)$. The two-qubit gate decomposition in terms of the two-qubit native gate is given by $YZ^{ij}(\theta_m) \equiv R_z^i(-\pi/2)~ CR^{ji}(J_{ij}\theta_m)~ R_z^i(\pi/2)$ (refer Fig.~\ref{fig:nativegatecirc}(b)). We arrange these gates in parallel before the optimization such that the resultant circuit is as shallow as possible. We ran the optimization with this circuit on a local simulator and obtained the optimal parameters $\theta_{opt}$ as before. 

We show the resultant probability distribution in Fig.~\ref{fig:superconducting_expt}(a) with 8192 shots. We can achieve a ground state probability of around 20\%. Apart from choosing the best layout, we implemented several additional strategies like {\it SWAP} strategy for achieving the desired connectivity~\cite{weidenfeller2022scaling}, native gate and pulse optimization strategies~\cite{Earnest_2021}, dynamical decoupling~\cite{Farfurnik2015}, and finally measurement error mitigation techniques~\cite{Nation2021}. Detailed discussions about these techniques are given in Appendix~\ref{ibmqappex}. After the circuit optimization and error mitigation, the success probability is enhanced from around 8\% to 20\%. Fig.~\ref{fig:superconducting_expt}(a2) show the configuration of the protein. The nearest neighbor connection is between `A'--`F', and hence $q_{1,6}=1$. The optimal turn sequence is given by $t_{expt} = [\bar1, 0, \bar3, 1, \bar0, 1]$. This configuration is one of the two most stable (the other one is shown in Fig.~\ref{fig:superconducting_expt}(b2)) configurations. Scaling to higher qubits for IBM devices is a major challenge due to the increased requirement of {\it SWAP} gates and finite coherence times of the qubits. Hence, CD-inspired ansatz is hardware-implementable but still requires improvements to make it scalable for hardware with low connectivity.  

\subsubsection{Google QVM} We implemented a 7 amino acid protein APRLRFY ($N=9$) with $p=1$ layers of the ansatz on quantum virtual machine \textit{rainbow}. Since we opted for a 2D grid connectivity, we selected one out of three possible $3\times 3$ grids with low error rates to map our problem. The native entangling gate $CZ$ was selected and $YZ(\theta)$ interactions were implemented by first creating $ZZ^{ij}(\theta_m) \equiv H^j CZ^{ij}H^j~ R_z^j(J_{ij}\theta_m)~ H^j CZ^{ij}H^j$. From here, $YZ^{ij}(\theta_m) \equiv R_x(\pi/2) ZZ^{ij}(\theta_m) R_x(-\pi/2) $ (Fig.~\ref{fig:nativegatecirc}(c)). In this implementation, $N_{2loc}=25$ two-body interactions were present in the Hamiltonian. 

After performing the circuit optimization (Appendix~\ref{ibmqappex}), the probability distribution of the ground state with 10000 shots is shown in Fig.~\ref{fig:superconducting_expt}(b). Around 48\% success probability was achieved with the optimized circuit. This behavior is mainly due to the circuit decomposition and the connectivity of the system. Also, as this is a noisy emulator, the real hardware is expected to show a lower success probability. The folded protein is shown in Fig.~\ref{fig:superconducting_expt}(b2). The nearest neighbor connection is between `A'--`F' which is shown by a green dotted line which means that interaction qubit $q_{1,6}=1$. The optimal turn sequence is given by $t_{expt} = [\bar1, 0, \bar3, 1, \bar0, 2]$ . We can observe that as compared to the previous case, the change in this sequence is only the last turn. Both these cases are optimal and thus the Hamiltonian we considered for APRLRFY protein has a doubly degenerate ground state.

\section*{Discussions} \label{Discussions}
We proposed a hybrid digitized-counterdiabatic quantum algorithm to investigate the protein folding problem. The parameterized quantum circuit associated with this algorithm is inspired by counterdiabatic protocols, has $\mathcal{O}(N^2)$ parameterization, and consists of only one-qubit and two-qubit gates. We applied this algorithm to a tetrahedral lattice protein folding problem, encoded in a 5-local Ising Hamiltonian where the ground state contains the corresponding protein configuration. We study various proteins with increasing amino acid chains and show that the proposed algorithm has an excellent performance in terms of convergence and circuit depth. These results outperform previously utilized quantum algorithms, enhancing the experimental feasibility with various circuit optimization strategies. We prove that claim by implementing up to 9 amino acid proteins with 17 qubits on several quantum hardware like trapped-ions and superconducting systems and achieving high success probabilities.

This work paves the way for implementing problem-inspired ansatz to industrial use cases in the current NISQ era by utilizing digitized counterdiabatic protocols. We believe this quantum algorithm can be extended to other relevant applications. There is also a connection between counterdiabatic protocols with adiabatic gauge potentials and dynamical Lie algebra. As mentioned before, some challenges need to be addressed, for instance, the sensitivity toward the initial parameters and the choice of the appropriate CD terms from the adiabatic gauge pool. Using various modifications like global optimizers would be interesting, as these algorithms find it difficult to reach the exact ground state at high system sizes. This is crucial as the protein structure changes significantly if we end up in an excited state. This also motivates the use of purely digitized-counterdiabatic quantum algorithms where the task is to find the exact ground state~\cite{hegade2021shortcuts}. We also emphasize that using digital-analog quantum computing with digitized-counterdiabatic quantum computing can also be crucial. We believe this work advances quantum computing, bringing us one step closer to practical quantum advantage, which would have to challenge success in classical computing including the recent AlphaFold achievements~\cite{alphafold}.

\section*{Methods}\label{Methods}
\subsection{Variational quantum algorithms and counterdiabaticity} \label{Variational quantum algorithms and counterdiabaticity}
In VQA, a circuit ansatz and a classical optimizer combine to solve an optimization problem. The task of the circuit ansatz is to generate quantum states based on the optimal parameters provided by the classical routines. The classical parts can be done by gradient-based optimizers (like Adam or Adagrad) or gradient-free optimizers (like COBYLA or POWELL). The goal is to minimize a cost function $C$ that can take various forms depending upon the problem. But in our case, we will use the expectation value of the problem Hamiltonian $H$ given by
\begin{equation}
    C(\vec{\theta}) = \bra{\psi(\vec{\theta})}H\ket{\psi(\vec{\theta})},
     \label{exp_ham}
\end{equation}
where $\vec{\theta} = \{\theta_1,\theta_2,...\}$ shows the parameters associated with the circuit ansatz. 
As mentioned earlier, both problem-inspired ansatz and hardware-efficient ansatz have their advantages and disadvantages. Despite the promising performance of the problem-inspired ansatz, the current hardware experience several bottlenecks like limited qubit connectivity, imperfect implementation of gates, limited coherence times, etc. This makes the implementation impractical, so that hardware-efficient ansatz is preferred. Generally, the hardware-efficient ansatz is of the form
\begin{equation}
    U(\vec{\theta}) = \prod_{k=1}^p U_k(\vec{\theta}_k) W_k ,\label{hea_form}
\end{equation}
where $\vec{\theta}_k$ are the optimizable parameters and $p$ is the number of layers. $U_k = exp[-i\theta_k V_k]$, where $V_k$ is a Hermitian operator. $W_k$ are non-parameterized gates usually consisting of 2-qubit connecting gates like $CNOT$s and $U_k$ are parameterized single qubit rotations. 

On the other hand, the problem-inspired ansatz use evolutions of the form
\begin{equation}
    U(\theta) = e^{-i \hat{g} t }\label{piaeq},
\end{equation}
where $\hat{g}$ is a Hermitian operator and $t$ is a parameter. These $\hat{g}$ are derived from the system of interest. For instance, in QAOA~\cite{farhi2014quantum}, $\hat{g}= H$ corresponds to the problem Hamiltonian, thus Eq.~\eqref{piaeq} resembles the trotterized-time evolution. In QAOA, the quantum circuit consists of two unitaries: Hamiltonian term $U_c(\gamma)$ and mixing term $U_b(\beta)$ applied $p$-times to the initial state $\ket{\psi_0}$. Here, $(\gamma,\beta)$ are the parameters to optimize by the classical optimizer. Along these lines, the evolution looks like
\begin{equation}
    \ket{\psi_f} = U_b{(\beta_p)}U_c{(\gamma_p)}U_b{(\beta_{p-1})}U_c{(\gamma_{p-1})} \dots U_b{(\beta_1)}U_c{(\gamma_1)}\ket{\psi_0}, \label{qaoa_form}
\end{equation}
where $\ket{\psi_f}$ shows the output state, while generally $U_b(\beta)= e^{-i\beta \sum_i \sigma_i^x}$ and $U_c(\gamma)=e^{-i\gamma H}$. Mixer term $U_b(\beta)$ can take several other forms as well~\cite{a12020034}. The aim is to minimize the cost function given by Eq.~\eqref{exp_ham}. QAOA directly relates to the quantum adiabatic evolution~\cite{Barends2016} hence it has believed to be a successful algorithm at large $p$ layers due to the adiabatic theorem. That being said, implementing circuits with large $p$ results in high circuit depths, not feasible for current near-term devices. Many adaptations to QAOA have been reported~\cite{a12020034,headley2020approximating,PhysRevResearch.4.033029}. Among them, CD protocol has been of interest recently~\cite{wurtz2022counterdiabaticity,yao2021reinforcement,chandarana2022digitized}, from which the newly proposed digitized-counterdiabatic QAOA (DC-QAOA) reports that the addition of CD terms to the usual QAOA ansatz gives significant improvements in the performance of QAOA. Finding CD terms is a critical task and is done by obtaining approximate CD terms by the adiabatic gauge potentials using the nested commutator (NC) method~\cite{PhysRevLett.123.090602}. In the considered NC method, the approximate CD terms are given by $ A_{\lambda}^{(l)}$ which can be calculated as
\begin{equation}
    A_{\lambda}^{(l)} = i \sum_{k = 1}^l \alpha_k(t) \underbrace{[H_{a},[H_{a},......[H_{a},}_{2k-1}\partial_{\lambda} H_{a}]]],
    \label{gauge}
\end{equation}
where $l$ shows the order of expansion and $H_{a}(\lambda) = ( 1 -\lambda(t)) H_{mixer} + \lambda(t)H$. The CD term is then digitized and the coefficient $\alpha$ is considered as an additional free parameter along with $(\beta,\gamma)$ to increase the ansatz expressibility. Summing up, DC-QAOA uses three unitary terms iteratively $p$-times: Hamiltonian term, mixer term, and CD term to minimize the cost function more effectively. However, we have to pay the price in terms of increased circuit depth per layer. To upgrade this method, we propose a CD-inspired circuit ansatz that is also hardware implementable. Hence, it partakes the advantages of both problem-inspired and hardware-efficient ansatz (see Results).

\subsection{Protein folding} \label{Protein folding}
\begin{figure*}[t]
    \centering
    \includegraphics[width=1\linewidth]{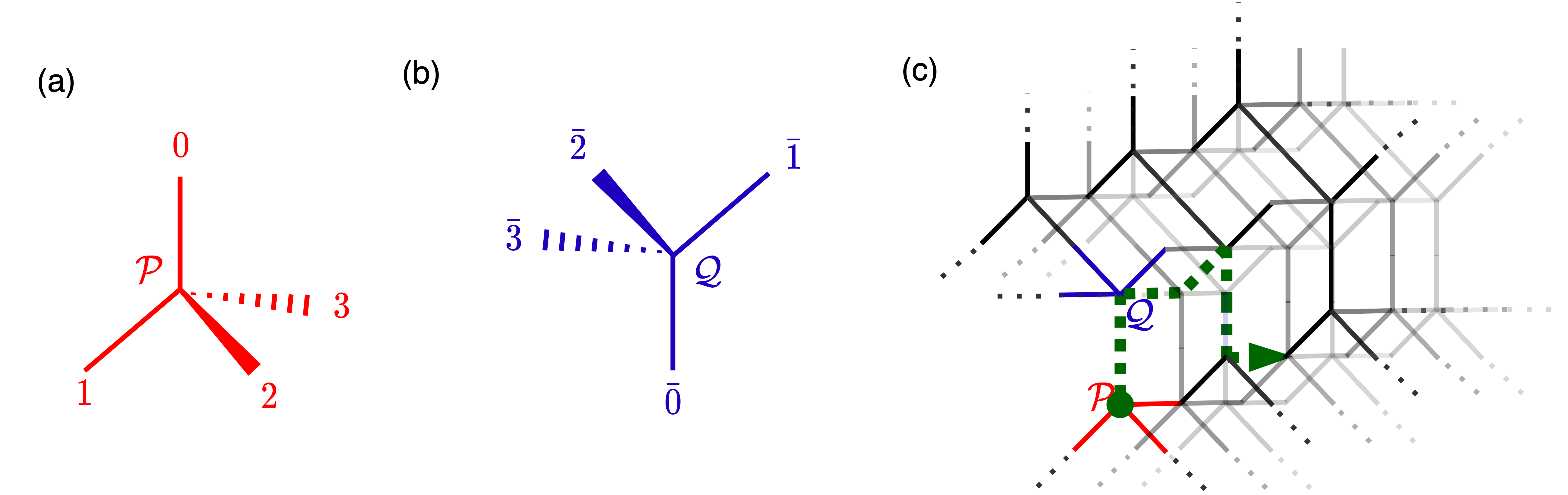}
    \caption{Schematic diagram showing the tetrahedral lattice structures. (a) shows the lattice $\mathcal{P}$ and (b) shows the inverted lattice $\mathcal{Q}$. The turns  taken by the amino acids will be one of the four directions ($t= 0,1,2,3$ or $ \bar0, \bar1, \bar2, \bar3$) and both lattices are switched at each turn. (c) shows the mesh made of $\mathcal{P}$ (shown in red) and $\mathcal{Q}$ (shown in blue), while the green line shows a schematic of the turns taken by an arbitrary protein in 3D.}
    \label{latfig}
\end{figure*}
In this section, we describe the current quantum scenario regarding protein folding, tackled as a lattice problem where amino acids are sequentially added to a given lattice such that the total conformation energy is minimum. The lattice can be 2D (plane) or 3D (cubic or tetrahedral). The complexity increases drastically while going from 2D to 3D because of the highly increasing number of possible protein configurations.
After selecting the lattice structure, the next important step is to choose the type of encoding. The widely studied encoding types are position encoding, where the lattice coordinates are encoded as qubits, and turn encoding, where the turn direction is encoded as qubits. In position encoding, the solution bitstring will represent the coordinates. Here, the respective amino acids should be placed for energy minimization, and, for turn encoding, the solution contains information about the turns taken by each amino acid sequentially to result in minimum energy configuration. The difficulty and the form of the Hamiltonian are highly dependent on the type of encoding chosen. Another vital task is to assign amino acids' interaction energies. There are two types of interactions, namely, the Hydrophobic model (HP), where the interaction coefficients are limited to binary values~\cite{PhysRevA.78.012320}, and the Miyazawa and Jernigan (MJ) interaction, where the coefficients are arbitrary depending upon the contact of amino acids~\cite{MIYAZAWA1996623}. Depending upon the structure, we have to define some constraints that will help us avoid configurations that are not allowed, like stacking different amino acids on the same lattice point.

In this work, we consider the 3D tetrahedral lattice model introduced in Ref.~\cite{Robert2021} with the turn encoding and MJ interactions. The qubits are encoded to represent the turn $t_i$ taken by the $(i+1)^{th}$ amino acid after $i^{th}$ amino acid. There are two sets of lattice $\mathcal{P}$ and $\mathcal{Q}$ exactly inverse to each other where the possible four turns are shown by $t=0,1,2,3$ for $\mathcal{P}$ (and $\bar{0},\bar{1},\bar{2},\bar{3}$ for $\mathcal{Q} )$ and each of the numbers shows the different direction of the tetrahedron shown in Fig.~\ref{latfig}. We devote two qubits to encode each turn so that the turns are given by $t_j=q_jq_{j+1}$ which scales as $2(N_a-3)$ where $N_a$ is the number of amino acids considered. $\mathcal{P}$ and $\mathcal{Q}$ are changed at every turn so that even turns can be shown by $\mathcal{P}$ and odd turns can be shown by $\mathcal{Q}$. An example of the turn bitstring will be of the form $t= [\bar1,0,\bar3,...]$. The schematic diagram is shown in Fig.~\ref{latfig}. The initial two turns can be fixed to $t_1=1$ and $t_2= \bar0$ due to the symmetry of space we consider. In addition, one more qubit can also be saved due to space symmetry if no side chains are considered. Hence, the conformation qubits $Q_c$ will look like
\begin{equation}
Q_c=[00][01][q_5 1][q_7 q_8].....[ q_{2(N-1)-1}q_{2(N-1)}] \, ,
\end{equation}
where $q_6=1$ due to spatial symmetry. To keep track of the turns, a function $g_m$ where $m={0,1,2,3}$ is constructed and this function will return 1 if the axis $m$ is selected at $i^{th}$ turn. The shortest distance between any two beads can be found by keeping track of the number of turns the beads had taken with lattices $\mathcal{P}$ and $\mathcal{Q}$. As far as the constraints are concerned, there are two types, namely, growth constraints that penalize unwanted conformations and chirality constraints that enforce correct values. To impose these, two terms $H_{gc}(Q_c)$ and $H_{ch}(Q_c)$ are added to the problem Hamiltonian with positive Lagrange multipliers $(\theta_{gc}, \theta_{ch})$. Details about how to create these functions are given in Ref.~\cite{Robert2021}. Along with $Q_c$, a set of qubits $Q_{in}$ are included that takes account of the interactions between the nearest-neighbor beads in the protein chain. $Q_{in}= q_{i,j}$ are a set of two-indices qubits that have information about nearest neighbor contact with $i^{th}$ and $j^{th}$ bead. If the contact occurs, the energy $e_{i,j}$ is applied to the Hamiltonian $H_{in}$. We consider the nearest neighbor interactions, where the contact is considered if the distance between two non-consecutive beads is unity. Thus the total qubits $Q_{tot} = \{Q_{c},Q_{in}\}$ and the  Hamiltonian $H$ is given by
\begin{equation}
    H(Q_{tot}) = H_{gc}(Q_c) + H_{ch}(Q_c) + H_{in}(Q_{in}).
\end{equation}
The conversion of $H(Q_{tot})$ into an Ising Hamiltonian $H$ will result in
\begin{equation}
\begin{aligned}
    H &= \sum_i h_i \sigma_z^i + \sum_{ij} J_{ij} \sigma_z^i \sigma_z^j +
    \sum_{ijk} K_{ijk} \sigma_z^i \sigma_z^j \sigma_z^k \\&
    + \sum_{ijkl} L_{ijkl} \sigma_z^i \sigma_z^j \sigma_z^k \sigma_z^l 
    + \sum_{ijklm} M_{ijklm} \sigma_z^i \sigma_z^j \sigma_z^k\sigma_z^l \sigma_z^m ,
\end{aligned}
\end{equation}
where the indices and the coefficients depend on specific proteins. Hence, $H$ is a 5-local Ising Hamiltonian whose ground state shows the optimal turn sequence for the folded protein. The specific reason for the selection of this model is that this model has a tetrahedral lattice that captures many physical and chemical properties. This model uses a lower number of qubits for a specified amino-acid chain as compared to other methods, but this comes at the cost of increasing the locality of the Hamiltonian. In Results, we show that this increased locality does not affect our CD-inspired ansatz, and we can get to the optimal solutions by using only 2-local terms in the PQC. This happens in contrast to QAOA, where the circuit ansatz will require 5-local terms, which makes it challenging to implement on a real device.

\begin{acknowledgments}
We acknowledge the Azure quantum credits program for providing access to the Quantinuum trapped-ions systems. P.C. acknowledges Mikel Garcia de Andoin and Martin Larocca for useful discussions. This work is supported by EU FET Open Grant  EPIQUS (899368),  QUANTEK project (KK-2021/00070), the Basque Government through Grant No. IT1470-22, the project grant PID2021-126273NB-I00 funded by MCIN/AEI/10.13039/501100011033 and by ``ERDF A way of making Europe" and ``ERDF Invest in your Future", NSFC (12075145), and STCSM (2019SHZDZX01-ZX04). X.C. acknowledges ayudas para contratos Ram\'on y Cajal–2015-2020 (RYC-2017-22482).
\end{acknowledgments}
\appendix
\section{Parameter Scaling}\label{parameterscale}
\begin{figure}
    \centering
    \includegraphics[width=1\linewidth]{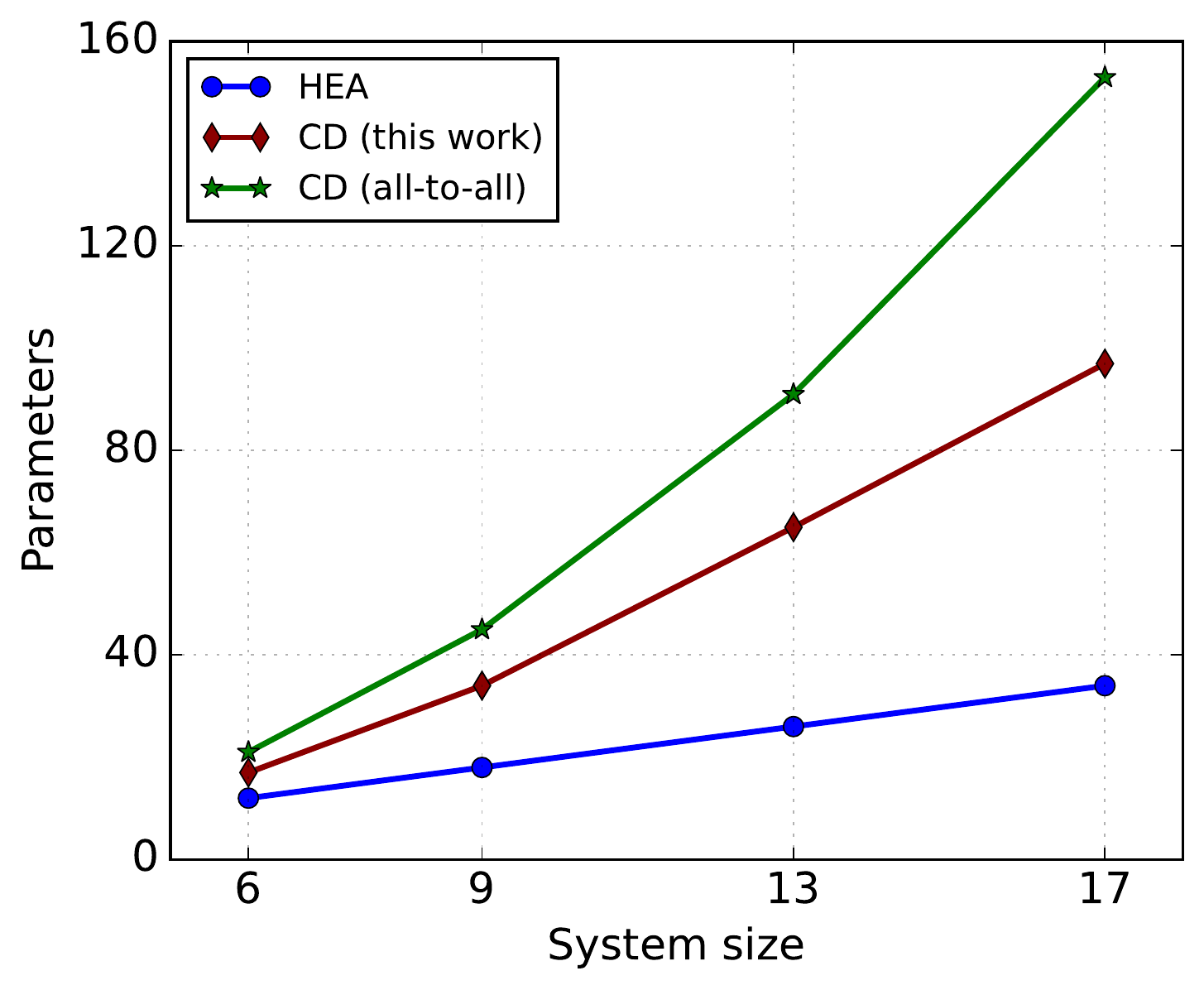}
    \caption{Number of optimizable parameters as a function of system size for various ansatz. The blue line shows hardware-efficient ansatz parameterization, the red line shows CD-inspired ansatz parameterization, and the green line shows the CD-inspired ansatz parameterization where all-to-all two-body interactions are present contrary to the protein folding problem. }
    \label{fig:new_Parameter_study}
\end{figure}

In Fig.~\ref{fig:new_Parameter_study}, we show how the parameters of $p=1$ CD-inspired ansatz of the protein folding problem vary as a function of system size as compared to both HEA and the case where all the possible two-body interactions are present.
\section{Native gate decomposition}
\begin{figure}
    \centering
    \includegraphics[width=1\linewidth]{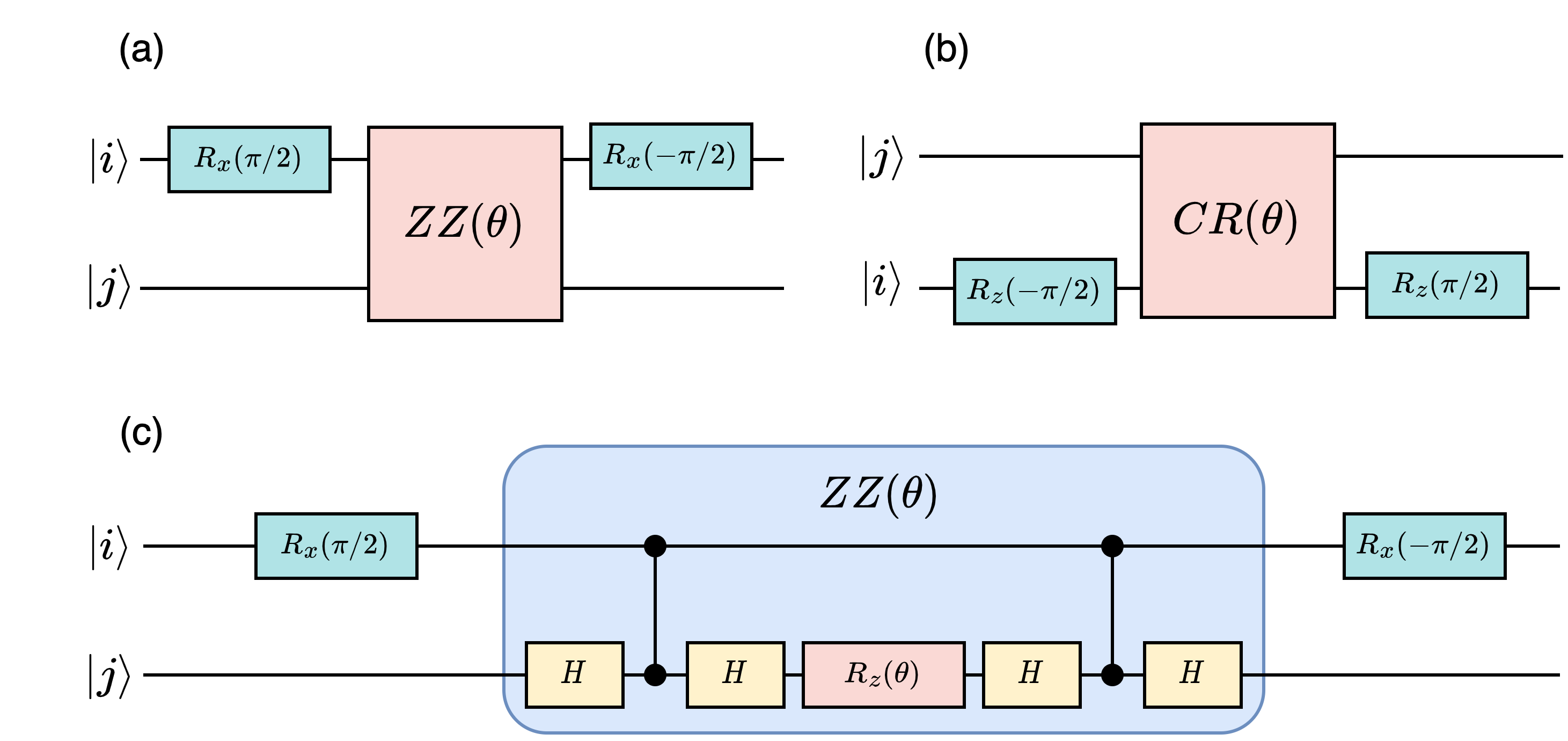}
    \caption{Native gate decomposition of $YZ(\theta)$ gate in real hardware for qubits $\ket{i}$ and $\ket{j}$. (a) shows the decomposition for Quantinuum tapped-ions hardware (native gate $ZZ(\theta)$), (b) shows the decomposition for IBM superconducting hardware (native gate $CR(\theta)$) and (c) shows the decomposition for Google QVM (native gate $CZ$). }
    \label{fig:nativegatecirc}
\end{figure}

The native gate decomposition of $YZ(\theta)$ interaction concerning various hardware is shown in Fig.~\ref{fig:nativegatecirc}. Fig.~\ref{fig:nativegatecirc}(a) shows the decomposition with respect to the $ZZ(\theta)$ interaction native to the Quantinuum trapped-ions hardware, Fig.~\ref{fig:nativegatecirc}(b) shows the decomposition in terms of $CR(\theta)$ gate native to IBM superconducting chip, and Fig.~\ref{fig:nativegatecirc}(c) show the decomposition in terms of $CZ$ gates native to the Google superconducting hardware.
\section{Experimental details}\label{ibmqappex}
\subsection{Quantinuum Trapped-ions}
\textbf{Hardware details.} The Quantinuum H1-1~\cite{quantinuum} is an all-to-all connected trapped-ion system with 20 qubits. There are multiple interaction zones where these physical qubits can move and the quantum operations are applied using lasers. Due to the existence of these zones, this device can perform parallel operations effectively. All-to-all connectivity is achieved by the physical rearrangement of qubits to perform all the two-qubit interactions. All the qubits are identical but the errors arise due to the quantum operations depending upon the locality of those qubits. Typical one-qubit gate infidelity is $4\times10^{-5}$, while the two-qubit gate infidelity is $3\times10^{-3}$ and the native gate-set are the two-qubit $ZZ(\theta)$ gates.

\textbf{Circuit optimization and Error mitigation.} Minimizing two-qubit gate error is an essential part of the experimental implementation. Due to all-to-all connectivity and suitable two-qubit native gate, we already have an implementable circuit. However, intending to obtain the best results, we strategically removed several $YZ(\theta)$ gates by analyzing the angles associated with them. Rotations with angles near zero do not contribute highly to the circuit but their gate errors do accumulate.  Thus these gates can be avoided to minimize two-qubit gate errors. This should be done carefully keeping in mind the error rates and change in fidelity. The final $ZZ(\theta)$ gate count at the time of implementation was reduced to 35 for $N=13$ qubits case and to 70 for $N=17$ qubits case.
\subsection{Google QVM}
\textbf{Hardware details.} The hardware from Google has Sycamore architecture with 53 transmon qubits arranged in a 2D grid. Hence, each physical qubit is connected to at most 4 other qubits. The single-qubit gates are executed by microwave pulses of a fixed frequency and two-qubit gates can be executed by bringing the nearest qubits on resonance and then turning on a coupling.  The two-qubit native gates are the $CZ$, $\sqrt{iSWAP}$, and Sycamore interactions. Typical two-qubit $\sqrt{iSWAP}$ implementation error is 1.4\% per gate when applied in parallel and the typical single-qubit gate error is 0.1\% per gate. More information can be found in Ref.~\cite{GoogleSupremacy}. In our experiment, we utilized the quantum virtual machine (QVM) offered by Google. Google offers two QVMs- $\textit{rainbow}$ and $\textit{weber}$- where a noise model is implemented that closely mimics the actual noise of the hardware. In this work, we utilize QVM $\textit{rainbow}$, a 23-qubit device with square-grid lattice connectivity (Fig.~\ref{fig:superconducting_expt}(b1)). 

\textbf{Circuit optimization and error mitigation.} As the connectivity is not all-to-all, a {\it SWAP} strategy is required to implement the circuit. We apply a {\it SWAP} strategy based on Ref.~\cite{weidenfeller2022scaling} that implements the circuit by using 11 {\it SWAP} gates as all the two-qubit interactions are not present in the circuit. In order to get the best results, we optimized the `moments' of the implemented circuit. A moment is defined as a set of operations, acting on different qubits such that all of them can be applied at a single abstract time slice. Essentially, the number of moments should be minimized to reduce the circuit execution time. In other words, we arranged the circuit such that the number of operations that can be applied at a single time is maximized. There are several ways to achieve this, for example, aligning the circuit to the left where the maximum possible number of operations are arranged from the start of the circuit. For our implementation, we optimized the circuit where all the operations are aligned into similar categories, that is, all the one-qubit and two-qubit operations are aligned in separate moments. This method is adopted to reduce the qubit idling since if these operations are isolated, the other qubits will be idle, leading to errors. Since we have only used the QVM, we limit ourselves to the circuit optimizations mentioned above. However, for the real device, there might be a need for further circuit optimizations. This can be done by adopting strategies like dropping moments that have a lower impact on the circuit results and adding dynamical decoupling to resist the qubit idling. Testing combinations of native gates to generate $YZ(\theta)$ interactions more effectively can also prove to be effective. For example, this is shown in Ref.~\cite{GoogleQAOA} where they use the Sycamore gates to generate the combination of $ZZ-SWAP$ interactions.
\subsection{IBM superconducting chips}
\begin{figure}
\centering
\includegraphics[width=1\linewidth]{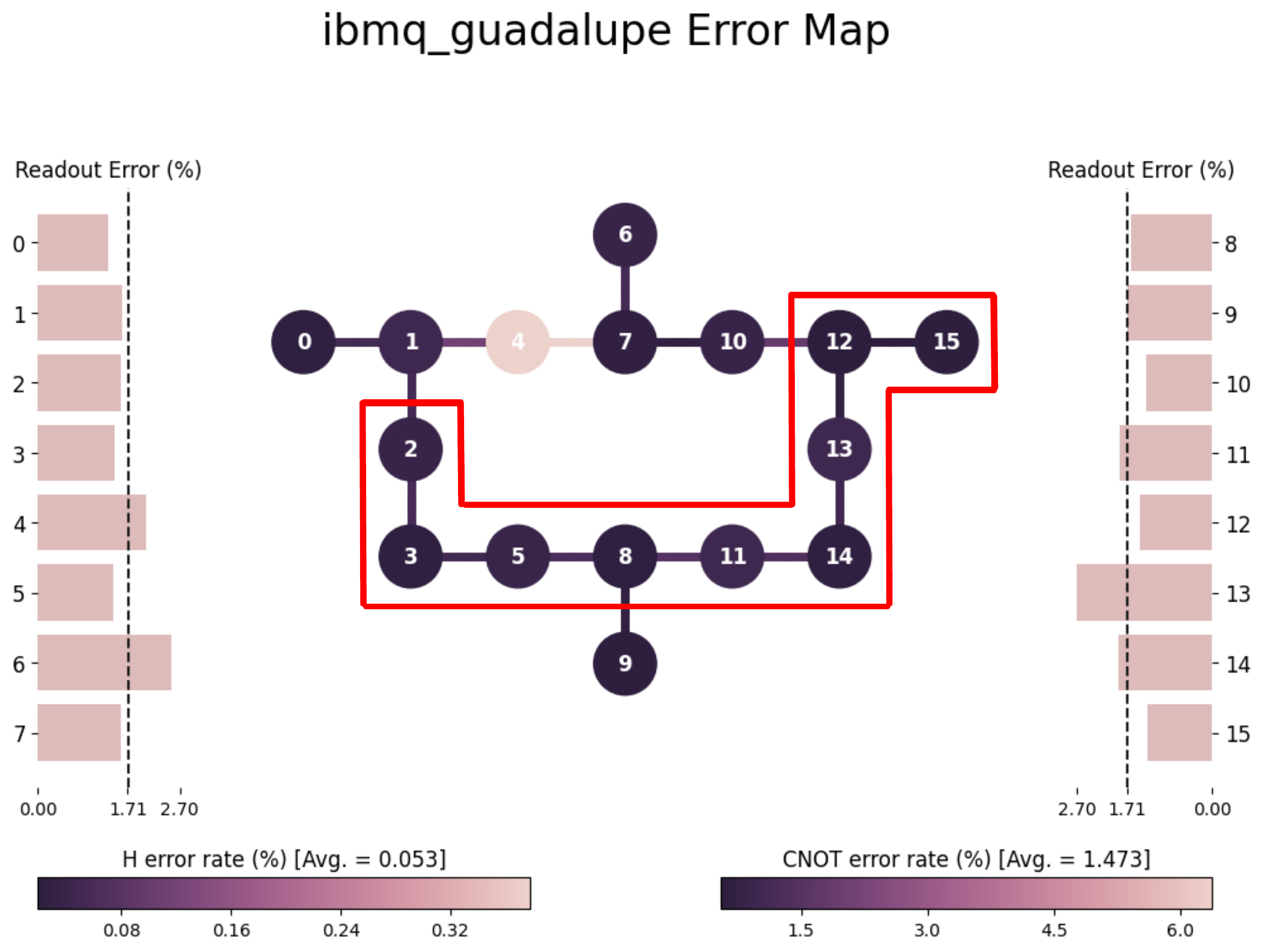}
\caption{Error map of IBM's 16 qubit \textit{ibmq\_guadalupe} device when the experiment was performed. The experiment was performed using the qubits highlighted in red color.}
\label{fig:guadalupe_device}
\end{figure}

\textbf{Hardware details.} IBM systems consist of fixed-frequency transmon qubits and the gate operations are applied with microwave pulses~\cite{PhysRevA.76.042319}. The connectivity is heavy-hex and for our implementation, we chose qubits with linear connectivity (Fig.~\ref{fig:superconducting_expt}(a1)). The two-qubit native gate for this device is the cross-resonance gate $R^{ij}_{zx}(\theta) = e^{-i \frac{\theta}{2} \sigma_i^z \sigma_j^x}$ also known as $CR(\theta)$ gate. The average single-qubit gate error is around 0.05\% and the average two-qubit $CNOT$ gate error is around 1.4\%. Error map of the $\textit{ibmq\_guadalupe}$ device when the experiment was performed is shown in Fig.~\ref{fig:guadalupe_device}. This error map shows the connectivity of the qubits, the Hadamard gate error of each of the qubits, the $CNOT$ error of all the present connections, and the read-out errors of the qubits. The qubits shown within the red box were selected for the experiment.

\textbf{Circuit optimization and error mitigation.}

\textit{ Best layout-}
Qubit selection is a crucial task while implementing the circuit as each qubit possesses its own single-qubit gate errors and two-qubit gate errors with the connecting qubits. To select the best qubits from the device, we implement a sub-graph isomorphism algorithm that selects 9 qubits out of 16 that minimize the expected gate errors~\cite{Nation2022}. 

\textit{\textit{SWAP} strategies-}
IBM's heavy-hex connectivity scheme means there only exists nearest neighbor coupling. As we have a linear connection of qubits, we need a good {\it SWAP} strategy to implement our circuit using the least {\it SWAP} gates possible. To achieve this, we implement the strategy performed in Ref.~\cite{weidenfeller2022scaling}. As some of the two-qubit interactions are missing, we could cover our connectivity by applying 30 {\it SWAP} gates for the case involving $N=9$ qubits. 

\textit{Native gate compilation-} {\it SWAP}, $CNOT$, and $ZZ(\theta)$ gates are not native to IBM's chips, being translated to the native $CR(\theta)$ gate before these circuits are executed. Ref.~\cite{Earnest_2021} showed how by being hardware aware, these native gates can be translated beforehand so by the annihilation of commuting gates we can reduce the depth of the final logical circuit. By being hardware-specific in this regime, we could also calibrate the $CR(\theta)$ gates so that no other translation is done. Therefore, the actual physical coupling represents the intended interaction gate between two qubits. Hence, the final pulse schedule gets reduced to its minimal expression according to the specifications of the hardware while preserving the intended structure for the ansatz. Once the circuit is reduced to its minimum expression, there are still some techniques we can apply for noise mitigation. Some are encoded within our pulse definition and others by statistically correcting the systematic error upon measurement.

\textit{Dynamical decoupling-} Dynamical decoupling techniques introduce different gate schemes to minimize the noise generated by the idling effect of the qubits while other longer-time two-qubit gates are being applied~\cite{PhysRevLett.121.220502}. They observed that if idling time can be used up by applying gates (that will not change the final result), the effect of noise can be reduced significantly. A well-known scheme is the $XY4$ scheme which applies a sequence of four rotation operations on the $X$ and $Y$ axis. Following a symmetrized version of the $XY4$ scheme as in Ref.  \cite{Farfurnik2015}, the $XY8$ scheme applies these rotations eight times instead of four.  For our experiment runs, we use the $XY8$ sequences shown by
\be
X_{\pi} \xrightarrow{} Y_{\pi} \xrightarrow{}...\xrightarrow{} X_{-\pi} \xrightarrow{} Y_{-\pi} .
\ee
Here, ${-\pi}$ and ${\pi}$ reflect the opposing pulse amplitude being applied on each axis. Thus, the result of the gate application renders the $I$ and by the active involvement, low-frequency noise is generated while idling gets reduced which results in an overall reduction in the noise. 

\textit{Measurement Error Mitigation-} Finally, readout error can be seen as systematically and consistent for a given chip, always representing a similar error distribution along the usage of a chip. By compensating for this final measurement error one could boost the final results to their statistically corrected error-less version. Usually, this is done by an approach where for a given ideally simulated probability distribution $\vec{p}_{ideal}$, upon the measurement on a real device would yield a noisy distribution $\vec{p}_{noisy}$ such that
\begin{equation}
   \vec{p}_{noisy} = \mathbf{A} \vec{p}_{ideal} . 
\end{equation}
Here, $\mathbf{A}$ represents a matrix called the assignment matrix that has information about the noise. The probability can be enhanced by finding $\mathbf{A^{-1}}$ and thus finding the error-mitigated quasi-probabilities. However, as the system sizes increase, the calculation of this matrix becomes difficult. In Ref. \cite{Nation2021}, an approach is proposed where, instead of calculating the full matrix, a reduced matrix is calculated. This is done by considering the qubits needed for the experiment runs instead of considering the full space, hence reducing the computational overhead. In our implementation, we use this technique to perform measurement error mitigation. 
\bibliography{reference.bib}
\end{document}